\documentclass[12pt]{article}
\usepackage[utf8]{inputenc}
\usepackage{times}
\usepackage{geometry}
\usepackage{graphics,graphicx}
\usepackage[small,sc]{caption}
\usepackage{amsmath}
\usepackage{subfigure}
\usepackage{epsfig}
\usepackage{wrapfig}
\usepackage{rawfonts}
\usepackage{pictex}
\usepackage{subfigure}
\usepackage{eucal}
\usepackage{amssymb}
\usepackage{setspace}
\usepackage[utf8]{inputenc}
\usepackage{enumitem,amssymb}
\usepackage{ragged2e}
\newlist{thematic}{itemize}{8}
\setlist[thematic]{label=$\square$}
\usepackage{pifont}

\date{}

\usepackage{graphicx}

\newcommand{\msun}{M_\odot}

\begin{document}

\begin{figure}[t!]
\includegraphics[width=1.0\textwidth]{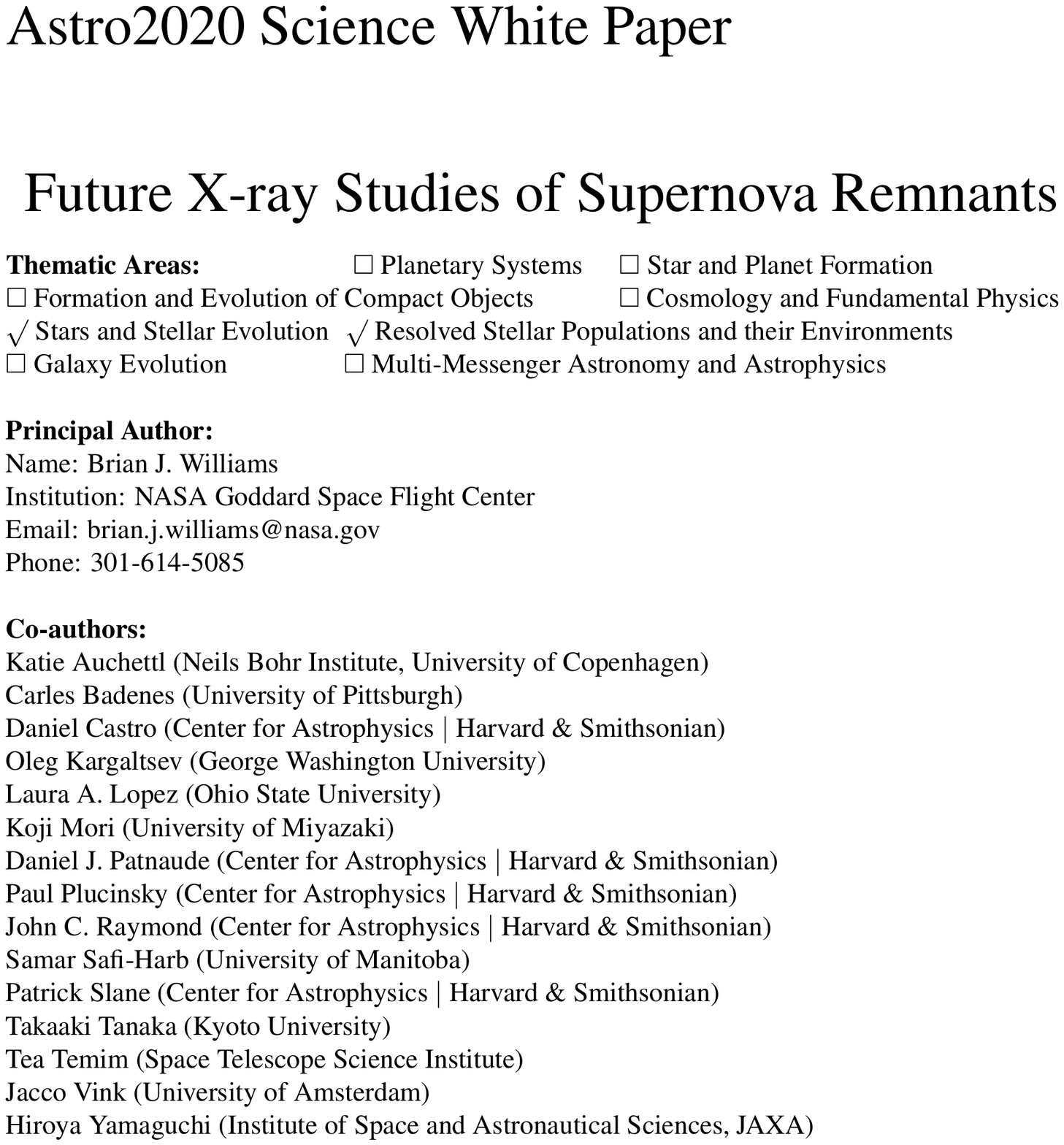}
\end{figure}

\newpage
\clearpage

\vspace{-24mm}
\section*{Introduction}
\vspace{-3mm}

\begin{wrapfigure}{R}{0.55\textwidth} 
\centering
\vspace{-4mm}
\includegraphics[scale=0.44]{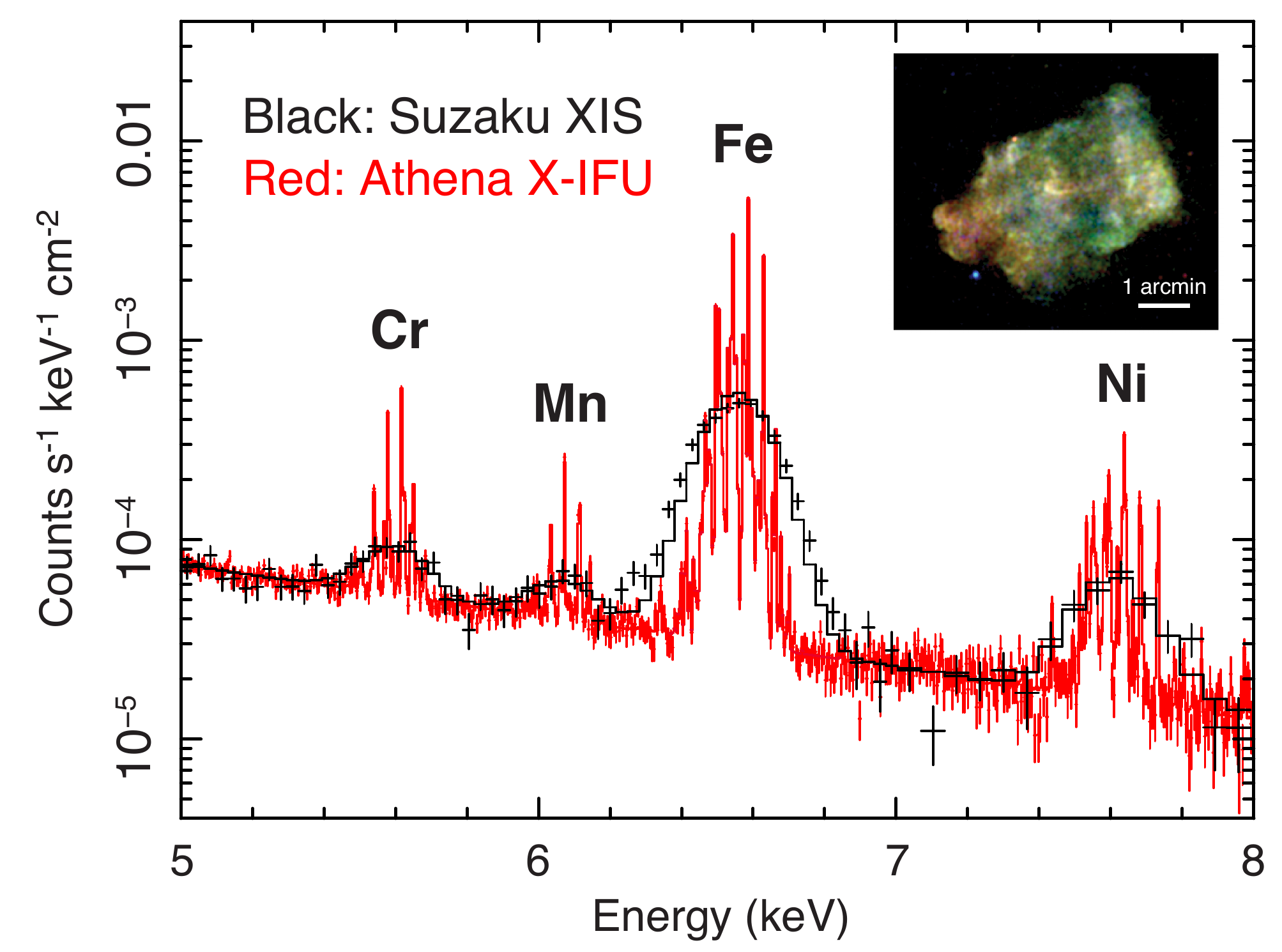}
\vspace{-4mm}
\caption{\footnotesize The spectra of the Fe-group elements in 3C 397 ({\it Chandra} image, inset), a several thousand year old galactic SNR that resulted from a Chandrasekhar-mass Type Ia SN \cite{yamaguchi15}. The observed Suzaku spectrum is plotted in black, with a simulated {\it Athena} spectrum plotted in red. 3C397 is $\sim 5'$ in diameter.}
\label{3c397}
\end{wrapfigure}

Few objects in the universe can rival supernova remnants (SNRs) for the multitude of astrophysical processes taking place, as well as their incredible morphological diversity at X-ray energies. Among other things, these objects are responsible for seeding the cosmos with heavy elements, accelerating particles up to extraordinary energies, and shaping the structure and chemical evolution of the interstellar medium (ISM) of galaxies. X-ray observatories like {\it Chandra}, {\it XMM-Newton}, and {\it NuSTAR} have made tremendous strides in studying SNRs from an imaging perspective and with low-resolution spectroscopy, but the higher-resolution dispersive spectrometers on these missions are not suited for extended objects like SNRs. {\em The next generation of X-ray telescopes will be ideally suited to study these objects more in depth.} For example, The X-ray Integral Field Unit (X-IFU) that will fly aboard {\it Athena} will have a resolution of 2.5 eV, and the notional calorimeter that would fly on Lynx would have a resolution of $0.3-3$ eV, depending on the array. The combination of spatial and spectral resolution will open a completely new region of parameter space, providing a critical complement to the thousands of unresolved distant SNe observed each year at optical, IR, and UV wavelengths. The power to disentangle position, velocity, and chemical composition in SN ejecta will be a defining benchmark for the next generation of multi-dimensional explosion models of both CC and Type Ia SNe, and hone in on the answers to several key open questions like the role of turbulence, the distribution of ignition spots, and the level of asymmetry in the ejected material. In this white paper, we explore several areas in which open questions regarding supernovae and their remnants can be addressed with X-ray studies in the next decade and beyond, focusing primarily on high-spectral resolution instruments. In companion Astro2020 white papers (Lopez et al. 2019, Safi-Harb et al. 2019), we emphasize the major advancements from the next generation of high-spatial resolution missions.

\section*{Supernova Explosion Physics}
\vspace{-3mm}

The era of high spatial resolution X-ray astronomy revolutionized SNR studies in many ways, among them by showing the spatial distribution of various ejecta products, from lighter elements like O, intermediate-mass elements (IMEs) like Si and S, and the heavier Fe-group elements (Fe, Cr, Mn, Ni). But this two-dimensional picture is fundamentally incomplete, and without the third dimension along the line of sight, it remains difficult to truly get a comprehensive picture of the ejecta distribution in young SNRs. For some objects, like Tycho?s SNR, efforts have been made to use CCD-quality spectral imaging to obtain information about the 3D structure \cite{williams17,sato17}, but uncertainties are substantial and the possibilities are limited to a few objects.

With the dawn of high-resolution spectroscopy, the third dimension is finally coming into focus. During a brief observation during the commissioning phase of the Hitomi mission in 2016, 17 photons from the Fe K$\alpha$ complex in N132D were sufficient to determine that the iron in the ejecta is moving with a bulk velocity of about 800 km s$^{-1}$ \cite{hitomi18a}. The short-lived {\it Hitomi} mission (along with its upcoming replacement X-ray Imaging and Spectroscopy Mission ({\it XRISM})) featured a 36-pixel microcalorimeter array with 30$''$ pixels; {\it Athena} will have arrays of thousands of pixels with sizes of order 1$''$ at the focus of a mirror with 5$''$ resolution and Lynx will have thousands of pixels varying in size from 0.5$''$ to 1.0$''$ with a 0.5$''$ mirror. With future missions, we will do spatially-resolved X-ray spectroscopy to measure the full three-dimensional morphologies of multiple ejecta products in various SNRs. By uncovering the spatial distribution and detailed spectral properties of, for instance, the Fe-group elements (see Figure~\ref{3c397}), we can directly compare SNR observations with hydrodynamic modeling of supernova explosions, filling in the gaps in our knowledge of the chain from stellar progenitor to supernova to SNR.

The nature of Type Ia SNe is an extremely important question, particularly given their importance as distance indicators for cosmological distances. We still do not have an agreed-upon model for how they explode.  The Fe-group elements have been shown to be a discriminator between various models of Type Ia SNe \cite{maeda10,seitenzahl15}. \cite{yamaguchi15} used deep Suzaku observations of 3C 397 to detect faint emission from Cr, Mn, and Ni, as shown in Figure~\ref{3c397}. They used the abundances of these metals to conclude that this remnant resulted from a white dwarf that exploded at the Chandrasekhar mass, perhaps implying a single-degenerate progenitor. 

These Fe-group element lines are too faint to map out the spatial distribution with {\it Chandra} or {\it XMM}. With a larger telescope, we could perform a census of these elements, along with the IMEs and the unburnt C and O, providing powerful constraints on the efficiency of thermonuclear burning and the density of the outermost layers of the white dwarf at the time of the explosion.

\section*{Relavistic Winds}
\vspace{-3mm}

For some core-collapse SNRs, a central pulsar injects a wind into the remnant interior and forms a high-pressure bubble, confined by the innermost slow-moving ejecta. The pulsar-wind nebula (PWN) expansion drives a shock into these ejecta, heating them and producing thermal emission. The shocks in young PWNe may be sufficiently fast to heat the surrounding ejecta to X-ray emitting temperatures, providing direct information on the density, composition, and velocity of the ejecta. At later stages of evolution, the SNR reverse shock encounters the PWN, temporarily halting its expansion. The PWN may be significantly disrupted during this interaction, leaving a relic nebula along with freshly injected particles in the immediate vicinity of the pulsar. 

Because most pulsars are born with significant kick velocities, they eventually escape the confines of their host SNRs. The rapid pulsar motion forms a bow-shock, with the pulsar wind swept back into a cometary tail. These systems are generally faint, requiring high sensitivity to probe the X-ray structure. 

\begin{figure}
\center
\includegraphics[width=1.0\textwidth]{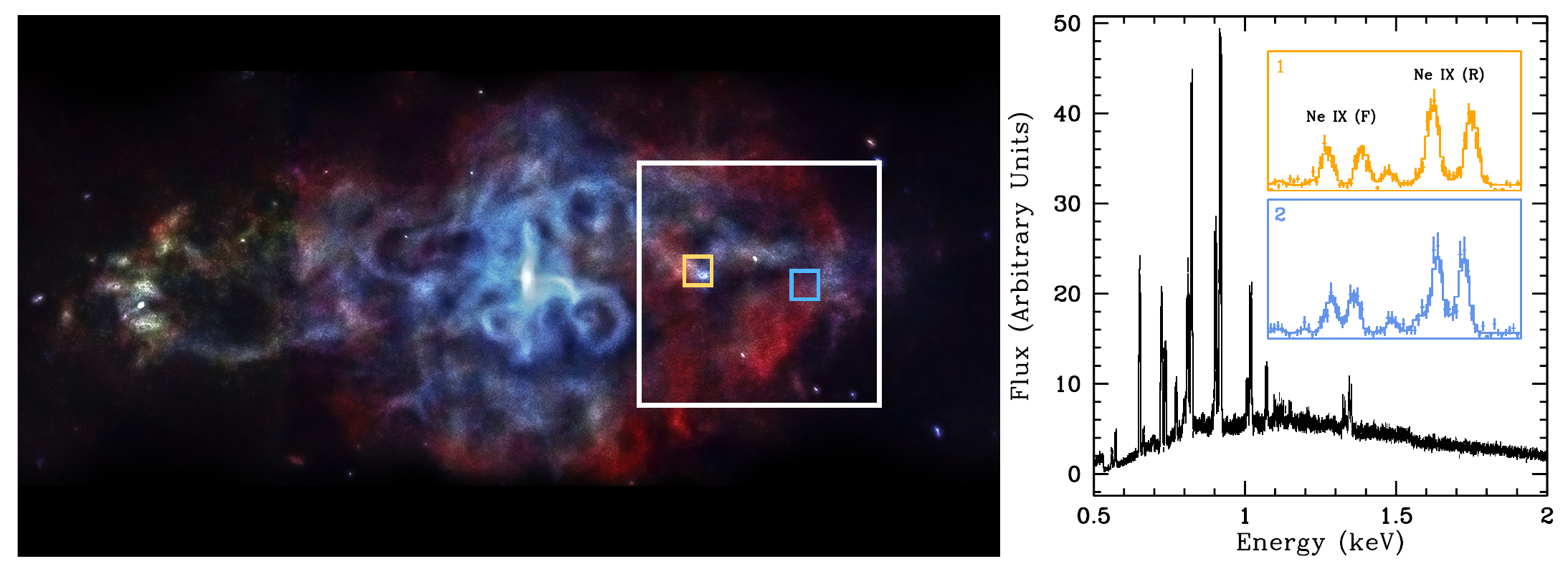} 
\caption{\footnotesize Left:  \textit{Chandra} image of 3C~58 with a $2.5^{\prime} \times 2.5^{\prime}$  spectral extraction region (white) and two $15^{\prime\prime} \times 15^{\prime\prime}$ extraction regions indicated. Right: Simulated high-resolution spectra from the entire region (black) and the Ne IX triplet from the two smaller regions (gold and blue insets). The expansion velocity of the shell is 900 km s$ ^{-1}$, as observed in optical filaments.  The Doppler-shifted features from the front and back shells are clearly separated, and the variation in the projected velocity with radius is also evident as a smaller separation between the lines in the western-most (blue) region.}
\label{3c58}
\vspace{-4mm}
\end{figure}

The weakness (or lack) of thermal X-ray emission from the youngest PWNe is puzzling, but holds important clues on the SN progenitor and the environment. This has been generally interpreted as due to a low-energy explosion and/or a low-density medium (see e.g., \cite{hitomi18b} for the Crab nebula and \cite{guest19} for G21.5--0.9). 3C 58 joins the handful of PWNe that lack thermal X-ray emission from an outer blast wave, but where a thin shell of thermal X-rays with an unknown origin surrounds the PWN (see Figure \ref{3c58}). The nature of this emission has important implications for constraining the SN progenitor of 3C 58; a CSM origin would imply that the PWN has already swept up most of the SN ejecta \cite{yang15}. An important next step would be to confirm whether the abundances are consistent with ejecta emission and resolve the kinematics of the material (see Figure \ref{3c58}).

PWNe are nature's most powerful accelerators, forming the majority of Galactic TeV sources detected by H.E.S.S. \cite{HESS18a,HESS18b}. Currently, the TeV and radio bands have been important for probing the evolutionary phase of the class of relic/older PWNe that occurs after the reverse shock interacts with the SN ejecta. Taking the next leap in our understanding of these objects will require X-ray missions with a substantial increase in sensitivity, high spatial and/or spectral resolution, and low instrumental background, which will enable the detection of these and other $\gamma$-ray bright, X-ray faint PWNe. These X-ray nebulae act as pathfinders for pulsar discovery or as tracers for confirming associations and the nature of many of the unidentified TeV sources. 

PWNe associated with supersonically moving pulsars are typically found around pulsars that left their host SNRs and are moving in the ISM. The ram pressure of the oncoming ISM diverts the shocked pulsar wind and confines it in an elongated structure (pulsar tail) behind the moving pulsar.  The tails, filled with ultra-relativistic particles emitting synchrotron radiation, are typically found in X-rays and radio, and are known to extend for several parsecs \cite{kargaltsev17}. A useful diagnostic of these phenomena can be obtained in X-rays by measuring the lengths and shapes of the extended (arcminute-scale) outflows and their spatially-resolved spectra \cite{pavan11,klingler16}. These X-rays are faint, and further study will require a combination of high sensitivity, large FOV, and low instrumental background. 

\section*{Shock Waves and Non-thermal Emission}
\vspace{-3mm}
Shock waves are ubiquitous in astrophysics. They convert kinetic energy to thermal energy in many environments and produce much of the X-ray emission observed.  SNR shocks are uniquely important because of their role in accelerating cosmic rays (they are thought to be the main source of cosmic ray protons below 3x10$^{15}$ eV), and because they offer the means to study faster (higher Mach number) shocks than are available for study in the solar wind. SNR shocks are collisionless, and as such there is no reason to expect them to produce equal temperatures among different particle species, or indeed to produce a thermal velocity distribution at all. Observations show that relatively slow shocks (350 km s$^{-1}$ in the Cygnus Loop) produce nearly equal proton, ion and electron temperatures, while faster shocks (2500-3000 km s$^{-1}$ in SN 1006) produce T$_{e}$/T$_{i}$ $<$ 10\% and roughly mass-proportional ion temperatures \cite{ghavamian13,raymond17,miceli19}. So far, only a handful of SNR shocks have been observed well enough to determine the temperatures of more than one species.

While shock waves are known to accelerate particles, the origin of cosmic rays is only partly understood. We know that turbulent magnetic fields accelerate electrons to $>$10 TeV, as evidenced by the X-ray synchrotron emission from virtually all young ($< 2000$ yr) SNRs. Moreover, the concentration of X-ray synchrotron emission from thin filaments can only be explained if electrons lose energy quickly, implying magnetic strengths of $>$ 100 $\mu$G, exceeding the compressed ISM field by an order of magnitude and providing evidence for magnetic field amplification by the streaming of cosmic rays. 

A high spatial resolution instrument that is an order of magnitude more sensitive will greatly advance our knowledge of cosmic-ray acceleration. Theory predicts that the particles will spend time upstream and downstream of the shock, yet no X-ray synchrotron emission has yet been found from this shock precursor. Finding this fainter component would directly measure the diffusion coefficient upstream, which determines how fast particles are accelerated and the level of magnetic field turbulence. 

An instrument with high sensitivity and high spectral resolution for extended sources is needed to advance our understanding of the wave-particle interactions that transfer energy among the different species. It should be able to make accurate measurements of the electron temperature from the shape of the bremsstrahlung continuum and from the relative intensities of emission lines such as the Fe K$\alpha$ to K$\beta$ ratio \cite{yamaguchi14}, as well as the ion temperature from the widths of the various spectral lines. 

Our view of cosmic rays in the local ISM concerns mostly those with energies above 100 MeV. However, a lot of energy may be contained in lower energy cosmic rays which diffuse more slowly. One way to detect these low energy cosmic rays is by detecting Fe K$\alpha$ emission, produced when $\sim 10$ MeV protons or $\sim 10$ keV electrons ionize the Fe K-shell, followed by fluorescence. Another way is to detect non-thermal bremsstrahlung from protons or electrons in the hard X-ray band as demonstrated by \cite{tanaka18}, who analyzed NuSTAR data of W49B. Based on the equivalent width of the Fe K$\alpha$ line with respect to the non-thermal bremsstrahlung continuum, we can determine if radiating particles are protons or electrons. A future X-ray mission, like {\it NuSTAR}, but with higher angular resolution in the hard X-ray band ($\sim 5 - 80$ kev) would allow this to be done for many more small, young remnants. A closely related problem is the longstanding difficulty in determining whether gamma-rays are produced by energetic protons or energetic electrons (hadronic vs. leptonic scenarios). Addition of the aforementioned ``new" spectral features will help us solve the problem as well. Detection of the Fe K$\alpha$ emission lines requires a spatial resolution of a few arcseconds, high spectral resolution, and an effective area many times that of Chandra. Detection of the non-thermal bremsstrahlung, on the other hand, requires a good sensitivity in the hard X-ray band above 10 keV.

\section*{Extragalactic Supernova Science}
\vspace{-3mm}

\begin{wrapfigure}{R}{0.55\textwidth} 
\centering
\vspace{-6mm}
\includegraphics[scale=0.46]{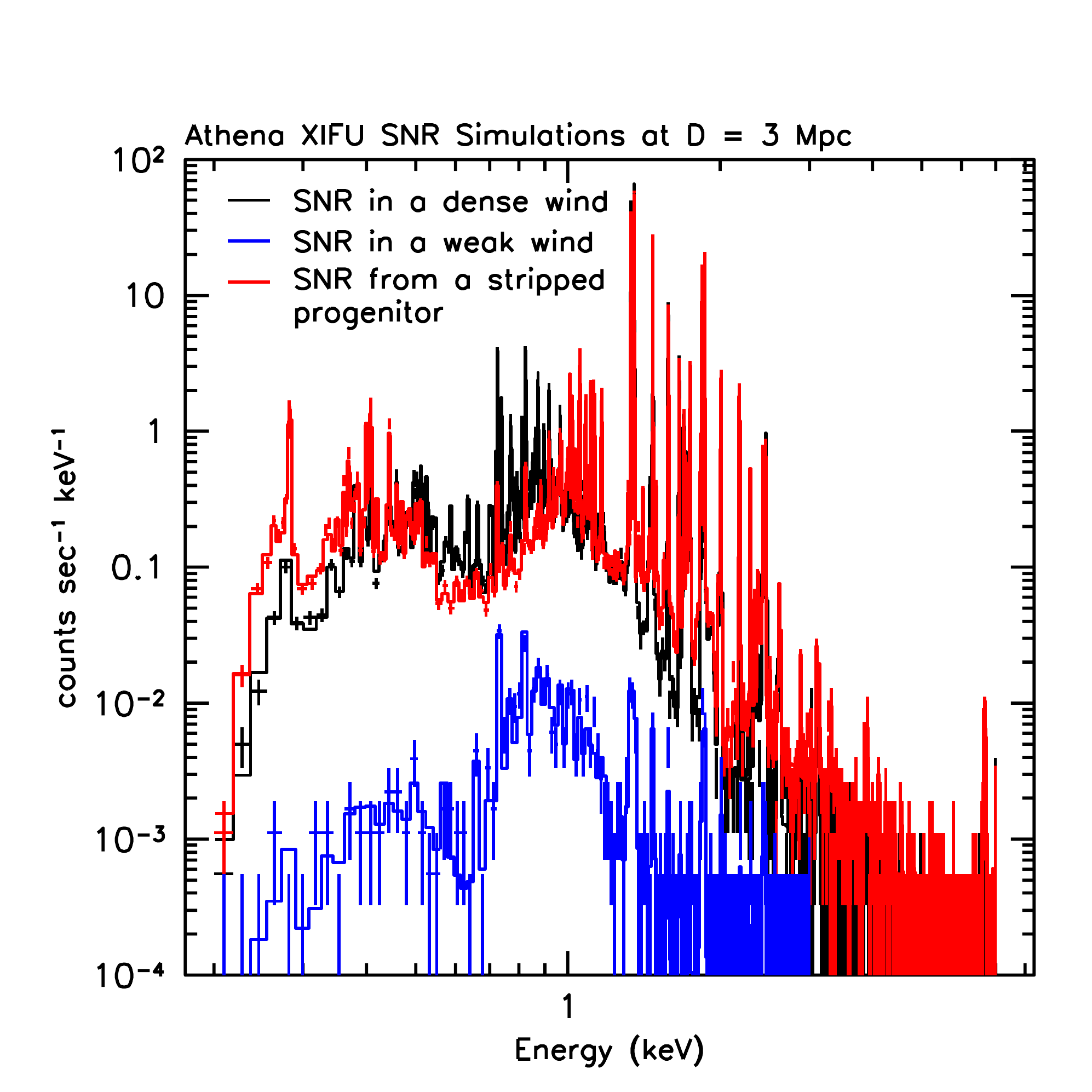}
\vspace{-4mm}
\caption{\footnotesize Athena X-IFU 100 ks simulations of three different SNRs at an age of ~500 yrs. Black and blue curves are IIP SNR with a dense wind (10 km s$^{-1}$ and 2 $\times 10^{-5}$ $\msun$ yr$^{-1}$) and a weak wind (20 km s$^{-1}$ and 1 $\times 10^{-5}$ $\msun$ yr$^{-1}$), respectively. Both have 8 $\msun$ of ejecta. The red is a stripped envelope SN/SNR, such as a Type IIb. It has 5 $\msun$ of ejecta and a higher explosion energy (2$\times 10^{51}$ ergs vs. 1$\times 10^{51}$).}
\label{3mpc}
\end{wrapfigure}

Population studies of SNRs in external galaxies provide insight into stellar lifecycles, galactic chemical evolution, and the star formation history of the host galaxy. Current observations of extragalactic SNRs are limited in scope. For example, the 700 ks Chandra mosaic of M83 \cite{long14} classified 87 of the 378 detected point sources as SNRs (compared with 225 optically-known remnants). Most of these did not have sufficient S/N for an X-ray spectrum. Observations of SNRs ranging from a few years to thousands of years can provide diagnostics on the circumstellar density surrounding the SN and the mass loss properties of the progenitor system, along with the SN ejecta density profile. Large numbers of well-measured radii and X-ray fluxes can be used, together with models for SNR populations, to constrain key open questions in stellar evolution, like the extent and frequency of CSM interaction in both Type Ia and CC SNe \cite{patnaude17,sarbadhicary18}; see Figure~\ref{3mpc}.

Combined with prior (non-)detections from {\it Chandra}, {\it XMM-Newton}, {\it Swift}, and even {\it ROSAT}, X-ray lightcurves can be constructed which provide a measure of the mass loss properties of the progenitor over 10$^{5}$ years. Chandra's detection limit of 10$^{-14.5}$ erg cm$^{2}$ s$^{-1}$ can only detect the most extreme mass-loss (2 $\times 10^{-5}$ $\msun$ yr$^{-1}$) at distances of $<$ 10 Mpc and ages of $<$ 1000 days. With the ability to probe an order of magnitude lower in mass loss and higher in age, we could observe the nebular phase of SNe from less extreme progenitors as they transition into an ejecta-dominated remnant. 

While core-collapse SNe are often X-ray bright at early stages, there are still only upper limits on the X-ray flux from Type Ia SNe (though see \cite{bochenek18} for a possible detection). The increased effective area of future large X-ray missions will significantly increase the likelihood of finding circumstellar interaction from SNe Ia in the first few years to decades. Another intriguing possibility is the as-yet-undetected $^{55}$Fe $\rightarrow$ $^{55}$Mn 5.9 keV radioactive decay line, which has been shown to be a discriminator between models of SNe Ia \cite{seitenzahl15}.  

While current X-ray missions have allowed great strides in our understanding of the explosive deaths of stars, the field is rife with possibilities for future growth. The next generation of X-ray telescopes must rise to this challenge with appropriate capabilities.

\end{document}